# Monte Carlo simulation of spin-reorientation transition in weak ferrimagnets YFeCrO3

E.V. Vasinovich[1], V.A. Ulitko[1], A.S. Moskvin[1,2]

[1]Ural Federal University, Ekaterinburg, Russia

[2]M.N. Mikheev Institute of Metal Physics, Ural Branch, Russian Academy of Sciences, Ekaterinburg, Russia

Abstract

This work presents the modeling of the magnetic $3d$-sublattice in mixed orthoferrites-orthochromites $YFe_{1-x}Cr_xO_3$ using classical Monte Carlo methods. It is shown that, when taking into account the competition of the Dzyaloshinskii vectors in the mixed compositions, magnetic moment compensations are observed, as well as angular magnetic configurations corresponding to the spin reorientation.

# 1 Introduction

Systems based on orthoferrites-orthochromites of the $RFe_{1-x}Cr_xO_3$ type (R = Nd, Gd, Dy, Y, Lu) were the subject of intensive fundamental theoretical and experimental research at the end of the 20th century [1-9]. A new surge of interest in these systems in the 21st century is associated with the discovery of specific magnetoelectric and magnetocaloric properties [10-13], as well as with the emerging prospects for the practical use of the phenomena of temperature compensation of the magnetic moment, exchange bias and spin reorientation for the creation of various multifunctional devices, for example, spintronics [14-22].

The fundamental research of orthoferrites and orthochromites primarily focuses on the features $4f - 3d$ of the interaction and the antisymmetric Dzyaloshinskii-Moriya (DM) exchange [1,23-27]. From the microscopic theory of antisymmetric exchange for systems of the $RFe_{1-x}Cr_xO_3$ type, one can obtain both a numerical estimate of the Dzyaloshinskii vector magnitude and, most importantly, its sign, which at one time played a fundamental role in the prediction and experimental discovery of a new type of magnetic ordering – weak ferrimagnetism [25]. The key feature of weak ferrimagnets is the competition of Dzyaloshinskii vectors between pairs of $Fe^{3+}$ - $Fe^{3+}$ and $Cr^{3+}$ - $Cr^{3+}$ ions on the one hand and pairs of $Fe^{3+}$ - $Cr^{3+}$ and $Cr^{3+}$ - $Fe^{3+}$ ions on the other hand.

Despite its long history, the unusual effect of spontaneous spin reorientation (SR) in weak ferrimagnets with a nonmagnetic R ion (Y, Lu) in the absence of external fields has not yet received an adequate description. Typically, researchers limit themselves to phenomenological approaches and mean-field approximations [1-3,25,28], which can be used to relate the possibility of a change in the Néel vector orientation $G_z \leftrightarrow G_x$ to the microscopic nature of the DM interaction in mixed orthoferrites-orthochromites $YFe_{1-x}Cr_xO_3$. Monte Carlo methods are tools that allow one to "observe" and study magnetic configurations in these systems. These methods have previously been used to model rare-earth perovskites [29-33] and, in particular, mixed orthoferrites-orthochromites [34-35], but the authors did not consider the spin reorientation phenomena.

Therefore, the main goal of the work is to develop a Monte Carlo (MC) method, which allows one to observe and study complex magnetic configurations in mixed orthoferrites-orthochromites $YFe_{1-x}Cr_xO_3$ that are atypical for the "parent" $YFeO_3$ and $YCrO_3$.

## 2 Model

Weak ferrimagnets of the YFe$_{1-x}$Cr$_x$O$_3$ type are orthorhombic perovskites with the space group *Pbnm*. There are 4 magnetic ions per unit $3d$ cell, for which the following classical basis vectors can be introduced [3]:

$$4S\boldsymbol{F} = \boldsymbol{S}^{(1)} + \boldsymbol{S}^{(2)} + \boldsymbol{S}^{(3)} + \boldsymbol{S}^{(4)},$$

$$4S\boldsymbol{G} = \boldsymbol{S}^{(1)} - \boldsymbol{S}^{(2)} + \boldsymbol{S}^{(3)} - \boldsymbol{S}^{(4)},$$

$$4S\boldsymbol{C} = \boldsymbol{S}^{(1)} + \boldsymbol{S}^{(2)} - \boldsymbol{S}^{(3)} - \boldsymbol{S}^{(4)},$$

$$4S\boldsymbol{A} = \boldsymbol{S}^{(1)} - \boldsymbol{S}^{(2)} - \boldsymbol{S}^{(3)} + \boldsymbol{S}^{(4)}. \tag{1}$$

Here, the vector $\boldsymbol{G}$ describes the fundamental antiferromagnetic component of the magnetic structure, $\boldsymbol{F}$ the vector of weak ferromagnetism (overt sublattice canting), while the weak antiferromagnetic components $\boldsymbol{C}$ and $\boldsymbol{A}$ describe the canting of the magnetic sublattices without the formation of a net magnetic moment (hidden sublattice canting). Typical spin configurations for the $3d$ sublattice, compatible with the antiferromagnetic sign of the fundamental isotropic superexchange, are denoted as $\Gamma_1(A_x, G_y, C_z)$, $\Gamma_2(F_x, C_y, G_z)$, $\Gamma_4(G_x, A_y, F_z)$, where the only nonzero components of the basis vectors appear in parentheses.

Unlike YFeO$_3$ and YCrO$_3$, which are weak ferromagnets with the main magnetic structure of the type $\Gamma_4$ $(G_x, A_y, F_z)$ below the Néel temperature $T_N$, the weak ferrimagnets orthoferrites-orthochromites YFe$_{1-x}$Cr$_x$O$_3$ According to magnetic measurements, complete or partial spin reorientation of the type is detected $G_xF_z - G_zF_x$ in a wide range of substitutions [2]. Typically, in such systems, reorientation occurs due to the $4f - 3d$ interaction [3], but in the case of the non-magnetic yttrium ion, such a mechanism is excluded and the anisotropy of the $3d$ sublattice must be considered. Indeed, the phenomenon can be explained by a strong decrease in the contribution of the DM interaction to the magnetic anisotropy [7,28].

Let us represent the spin Hamiltonian of a weak ferrimagnet in its simplest form, taking into account only the contributions of the isotropic exchange interaction, as well as the antisymmetric Dzyaloshinsky-Moriya exchange:

$$\hat{H} = \hat{H}_{ex} + \hat{H}_{DM},$$

$$\hat{H}_{ex} = \frac{1}{2} \sum_{\langle ij \rangle} I_{ij} (\hat{S}_i \cdot \hat{S}_j),$$

$$\hat{H}_{DM} = \frac{1}{2} \sum_{\langle ij \rangle} d_{ij} \cdot [\hat{S}_i \times \hat{S}_j], \qquad (2)$$

where the summation is over the nearest neighbors, $I_{ij}$ is the exchange integral, $d_{ij}$ and is the Dzyaloshinsky vector.

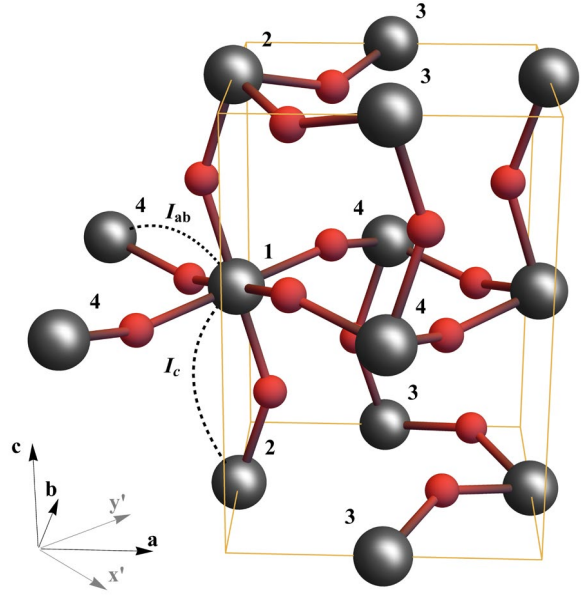

**Figure 1.** Structure of superexchange bonds; large balls are $Fe^{3+}$, $Cr^{3+}$ ions, small ones are $O^{2-}$; 1, 2, 3, 4 are magnetic ions in four nonequivalent positions

Figure 1 shows the structure of superexchange bonds in the model. The cation-anion distances and superexchange bond angles for nearest neighbors differ slightly, so below we assume the equality of the superexchange integrals $I_{ab} = I_c = I$ and the moduli of the Dzyaloshinsky vectors $d_{ab} = d_c = d$, although the vectors themselves point in different directions. We further assume that pairs of nearby ions lie along the axes of the coordinate system $x'y'z'$, which is rotated around the axis $z$ by an angle of 45 degrees; however, all vector quantities in this work are calculated in the system $xyz$, whose axes correspond to the axes of the crystal $abc$.

**Table 1.** Components $x$, $y$, $z$ of structure factors $[\mathbf{r}_i \times \mathbf{r}_j]$ calculated from neutron diffraction data [37] for YFe$_{0.5}$Cr$_{0.5}$O$_3$

|  | $x$ | $y$ | $z$ |
|---|---|---|---|
| $[\mathbf{r}_2 \times \mathbf{r}_1]$ | $\alpha_c = 0.216$ | $\beta_c = 0.562$ | 0 |
| $[\mathbf{r}_4 \times \mathbf{r}_1]$ | $\pm\alpha_{ab} = 0.303$ | $\beta_{ab} = 0.287$ | $\gamma_{ab} = 0.397$ |

The microscopic expression of the relationship between the Dzyaloshinskii vector and the geometry of the cation-anion-cation superexchange bond is [1]

$$\mathbf{d}_{ij} = d_{ij}(\theta) \cdot [\mathbf{r}_i \times \mathbf{r}_j], \qquad (3)$$

where $\mathbf{r}_{ij}$ are the unit vectors along the O$^{2-}$ – Fe$^{3+}$ or O$^{2-}$ – Cr$^{3+}$ bonds, $\theta$ is the superexchange coupling angle (will be omitted from the notation below). The structural factors determining the orientation of the Dzyaloshinskii vectors in orthoferrites-orthochromites of the YFe$_{1-x}$Cr$_x$O$_3$ type are given in Table 1. Simple formula (3) allows us to establish a direct relationship between magnetic noncollinearity (overt and hidden canting of sublattices) in weak ferromagnets with a crystalline structure [23-25].

In accordance with the symmetry of the crystal, the explicit form of the Dzyaloshinskii vectors depending on the site number is as follows:

$$\mathbf{d}_{ijk}(Ox') = d \cdot \begin{pmatrix} (-1)^{i+j+k}\alpha_{ab} \\ (-1)^{i+j+k}\beta_{ab} \\ (-1)^{i+j}\gamma_{ab} \end{pmatrix}, \qquad (4)$$

$$\boldsymbol{d}_{ijk}(Oy') = d \cdot \begin{pmatrix} -(-1)^{i+j+k}\alpha_{ab} \\ (-1)^{i+j+k}\beta_{ab} \\ (-1)^{i+j}\gamma_{ab} \end{pmatrix}, \tag{5}$$

$$\boldsymbol{d}_{ijk}(Oz') = d \cdot \begin{pmatrix} (-1)^{k}\alpha_{c} \\ (-1)^{i+j+k}\beta_{c} \\ 0 \end{pmatrix}, \tag{6}$$

where the index $i$ numbers the ions along the axis $x'$, $j$ along the axis $y'$, $k$ along the axis $z'$, $\boldsymbol{d}_{ijk}(Ox')$ is the vector for pairs of ions at the site $ijk$ of its nearest neighbor along the axis $x'$, $\boldsymbol{d}_{ijk}(Oy')$ is the vector for pairs of ions along the axis $y'$, $\boldsymbol{d}_{ijk}(Oz')$ is the vector for pairs of ions along the axis $z'$.

## 3 Methods

For the numerical simulation of a simple cubic $3d$ lattice with Hamiltonian (2), we considered two Monte Carlo (MC) methods with different ways of selecting states at lattice sites within the Metropolis algorithm [38].

In the first case, this is a pure classical Monte Carlo method (MC1), where the spin operators $\widehat{S}_i$ in Hamiltonian (2) are replaced by classical vectors

$$\boldsymbol{m}_i = S_i \cdot \begin{pmatrix} \cos v_i \cdot \sqrt{1 - u_i^2} \\ \sin v_i \cdot \sqrt{1 - u_i^2} \\ u_i \end{pmatrix}, \tag{7}$$

where $u_i$ is a random variable from -1 to +1, $v_i$ is a random variable from 0 to $2\pi$, $S_i$ and is the spin quantum number of the corresponding site. This is one of the simplest MC methods, which is often used when working with a spin Hamiltonian of type (2), including for describing magnetic phenomena in rare-earth perovskites [29,30,32,35,36].

At the initial step of MC, numbers are randomly selected $u_i^{(0)}, v_i^{(0)}$ at each lattice site $i$, and the magnetic moments $\boldsymbol{m}_i$ and energy of the system are calculated $E^{(0)}$. At the next step, new random numbers $u_i^{(1)}, v_i^{(1)}$ are selected, and a random number $p$ between 0 and 1 is selected, then the corresponding energy $E^{(1)}$ and its change relative to the previous state of the system are calculated $\Delta E = E^{(1)} - E^{(0)}$. If the new state satisfies the inequality $\exp(-\Delta E/T) > p$, where $T$ is the system's temperature, then we accept the new state of the system: $u_i^{(0)} \to u_i^{(1)}$, $v_i^{(0)} \to v_i^{(1)}$. If the new state does not satisfy the inequality, then we retain the previous state of the system. This step is repeated until the system reaches equilibrium at a given temperature $T$.

In the second case, we propose a semiclassical Monte Carlo (MC2) method, where the state of a site is given by the wave function $|\psi\rangle = \sum_M c_M |S, M\rangle$, where $S$ is the spin of the site, $M$ is the projection of the spin onto the z axis, $c_M$ are random coefficients with normalization $\langle \psi | \psi \rangle = 1$ (site indices are omitted).

In the case of iron ions (spin $S = 5/2$) the wave function has the form

$$|\psi\rangle = c_{-\frac{5}{2}} \left| \frac{5}{2}, -\frac{5}{2} \right\rangle + c_{-\frac{3}{2}} \left| \frac{5}{2}, -\frac{3}{2} \right\rangle + c_{-\frac{1}{2}} \left| \frac{5}{2}, -\frac{1}{2} \right\rangle$$

$$+ c_{\frac{1}{2}} \left| \frac{5}{2}, \frac{1}{2} \right\rangle + c_{\frac{3}{2}} \left| \frac{5}{2}, \frac{3}{2} \right\rangle + c_{\frac{5}{2}} \left| \frac{5}{2}, \frac{5}{2} \right\rangle, \qquad (8)$$

where the coefficients of the function can be represented as

$$c_{\frac{5}{2}} = \sqrt{1 - \zeta_1^{\frac{1}{5}}} e^{i 2\pi \xi_{\frac{5}{2}}},$$

$$c_{\frac{3}{2}} = \zeta_1^{\frac{1}{10}} \zeta_2^{\frac{1}{8}} \zeta_3^{\frac{1}{6}} \zeta_4^{\frac{1}{4}} \zeta_5^{\frac{1}{2}} e^{i 2\pi \xi_{\frac{3}{2}}},$$

$$c_{\frac{1}{2}} = \zeta_1^{\frac{1}{10}} \zeta_2^{\frac{1}{8}} \sqrt{1 - \zeta_3^{\frac{1}{3}}} e^{i 2\pi \xi_{\frac{1}{2}}},$$

$$c_{-\frac{1}{2}} = \zeta_1^{\frac{1}{10}} \zeta_2^{\frac{1}{8}} \zeta_3^{\frac{1}{6}} \sqrt{1 - \zeta_4^{\frac{1}{2}}} e^{i 2\pi \xi_{-\frac{1}{2}}},$$

$$c_{-\frac{3}{2}} = \zeta_1^{\frac{1}{10}} \zeta_2^{\frac{1}{8}} \zeta_3^{\frac{1}{6}} \zeta_4^{\frac{1}{4}} \sqrt{1 - \zeta_5} e^{i 2\pi \xi_{-\frac{3}{2}}},$$

$$c_{-\frac{5}{2}} = \zeta_1^{\frac{1}{10}}\sqrt{1-\zeta_2^{\frac{1}{4}}}e^{i2\pi\xi_{-\frac{5}{2}}}, \tag{9}$$

where all $\zeta_q$, $\xi_M$ are random variables between 0 and 1. In the case of chromium ions with spin $S = 3/2$, it is necessary to fix $\zeta_1 = \zeta_2 = 1$. This parameterization of the coefficients guarantees the normalization of the wave function at the site and that any state, given a uniform sample, $\zeta_q$, $\xi_M$, will appear equally frequently during the algorithm's execution.

Thus, in the MC2 method, the magnetic moment at a site and the energy of the system are calculated, respectively, as

$$\boldsymbol{m}_i = \langle\psi_i|\widehat{\boldsymbol{S}}_i|\psi_i\rangle, \tag{10}$$

$$E = \langle\Psi|\widehat{H}|\Psi\rangle, \tag{11}$$

where $|\Psi\rangle = \prod_i|\psi_i\rangle$ is the wave function of the entire system. This method operates using the same Metropolis algorithm as the one you provided for MC1, but uses formulas (8)-(11).

To characterize the angular phase at each step of the MC (after reaching equilibrium in the system), in both methods the basis vectors (1) are calculated on the entire lattice as

$$\boldsymbol{F} = \frac{1}{N}\sum_{ijk}^{N}\frac{\boldsymbol{m}_{ijk}}{S_{ijk}}, \quad \boldsymbol{G} = \frac{1}{N}\sum_{ijk}^{N}(-1)^{i+j+k}\frac{\boldsymbol{m}_{ijk}}{S_{ijk}},$$

$$\boldsymbol{C} = \frac{1}{N}\sum_{ijk}^{N}(-1)^{i+j}\frac{\boldsymbol{m}_{ijk}}{S_{ijk}}, \quad \boldsymbol{A} = \frac{1}{N}\sum_{ijk}^{N}(-1)^{k}\frac{\boldsymbol{m}_{ijk}}{S_{ijk}}, \tag{12}$$

where the index $i$ numbers the ions along the axis $x'$, $j$ along the axis $y'$, $k$ along the axis $z'$, $N$ is the number of $3d$ ions, $S_{ijk}$ is the spin of the site $ijk$, $\boldsymbol{m}_{ijk}$ are the magnetic moment of the site $ijk$, calculated using formula (7) in the MC1 method and formula (10) in the MC2 method.

## 4 Results

Numerical simulations were performed on a simple cubic lattice of $N = 64 \times 64 \times 64$ sites with periodic boundary conditions. Each site was randomly selected as an iron or chromium ion based on its concentration $x$. To establish equilibrium, at given temperature and concentration values, $2 \cdot 10^4$ MC steps were performed at each lattice site, followed by additional $3 \cdot 10^4$ MC steps per site to collect statistics.

Based on the results in the mean field approximation (MFA) [28], the Hamiltonian parameters in this work take the following values: $I_{FeFe} = 36.6K$, $I_{CrCr} = 18.7K$, $I_{CrFe} = 13.4K$, $d_{FeFe} = 2.0K$, $d_{CrCr} = 1.7K$, $d_{FeCr} = -2.5$ K, where the negative sign of the parameter $d_{FeCr}$, relative to $d_{FeFe}$ and $d_{CrCr}$, sets the competition of the Dzyaloshinsky vectors in the model system.

It should be noted that the $N = 64^3$ ion systems under consideration are far from the size of real samples, and the results may depend on how the impurity is distributed in a specific simulation. For example, with a significant amount of impurity ($x \approx 0.5$), the system will contain different regions: one where all the nearest neighbors of the selected ion are ions of the same type (where only the vectors $\boldsymbol{d}_{FeFe}$ and $\boldsymbol{d}_{CrCr}$ contribute), and another where all the nearest neighbors are ions of a different type (where only the vector $\boldsymbol{d}_{FeCr}$ contributes). Depending on the relative volume of the regions, the magnitude and direction of the basis vectors (1) will differ. We will not explore these possible effects in this paper.

Calculations using the MC1 method show that in the mixed orthoferrite-orthochromite YFe$_{1-x}$Cr$_x$O$_3$, below the critical temperature $T_N(x)$, a spontaneous transition occurs from a disordered state to a phase with only nonzero components of the basis vectors $G_x$, $A_y$, $F_z$, i.e., to the phase $\Gamma_4$. In this regard, the MC1 method is consistent with the MFA, including the conclusion that for a reorientation to a phase different from $\Gamma_4$ ($G_x, A_y, F_z$), competition between the Dzyaloshinskii vectors alone is insufficient and, for example, the effects of single-ion spin anisotropy must be taken into account.

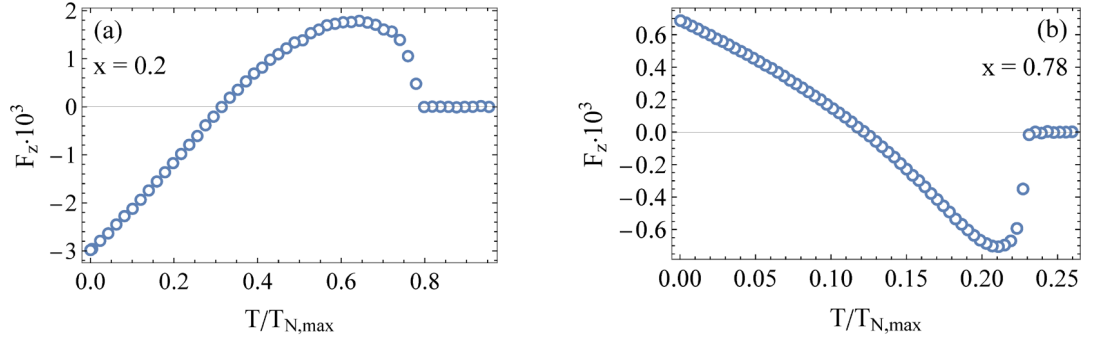

**Figure 2.** The first ( a ) and second ( b ) compensations of the weak ferromagnetic vector **F** in the MC1 method

Fig. 2 shows examples of the behavior of a weak ferromagnetic vector **F** depending on the relative temperature $T/T_{N,max}$, where $T_{N,max} = 320$ K is the temperature of formation of the magnetic moment in the YFeO$_3$ model system. Importantly, like MFA, the MC1 method shows the presence of compensation of the magnetic moment (the effect of negative magnetization) at a low concentration of chromium $x \approx 0.2$ (Fig. 2a) and also predicts the presence of a second compensation near the concentration $x \approx 0.8$ (Fig. 2b).

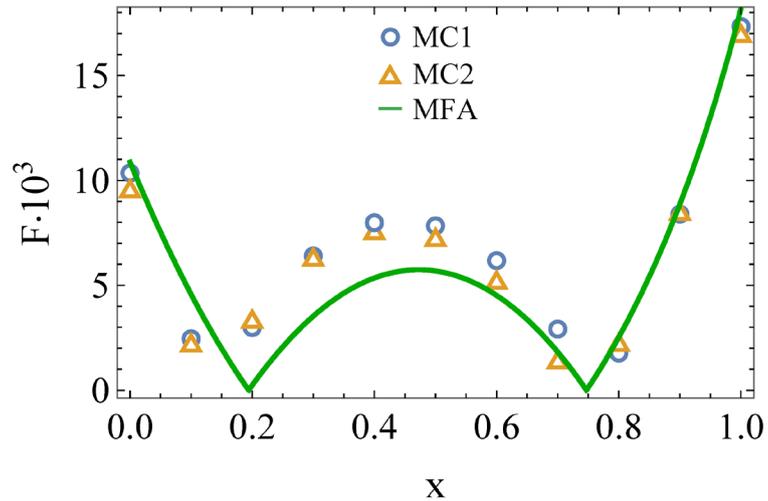

**Figure 3.** Concentration dependence of the weak ferromagnetic vector **F** in the MC1 (circles), MC2 (triangles) and mean field approximation (solid line) methods near the ground state ($T/T_{N,max} \ll 1$)

In the case of the MC2 method, for a number of concentrations, the temperature dependences of the absolute values of the vectors $F, G$ behave in the same way as in the case of the MC1 method, but with an ordering temperature $T_{N,max} = 85$ K. The magnetizations near the ground state are also close (Fig. 3). The first and second compensations, for example, at $x = 0.2$ and $x = 0.77$, respectively, are also observed here (Fig. 4).

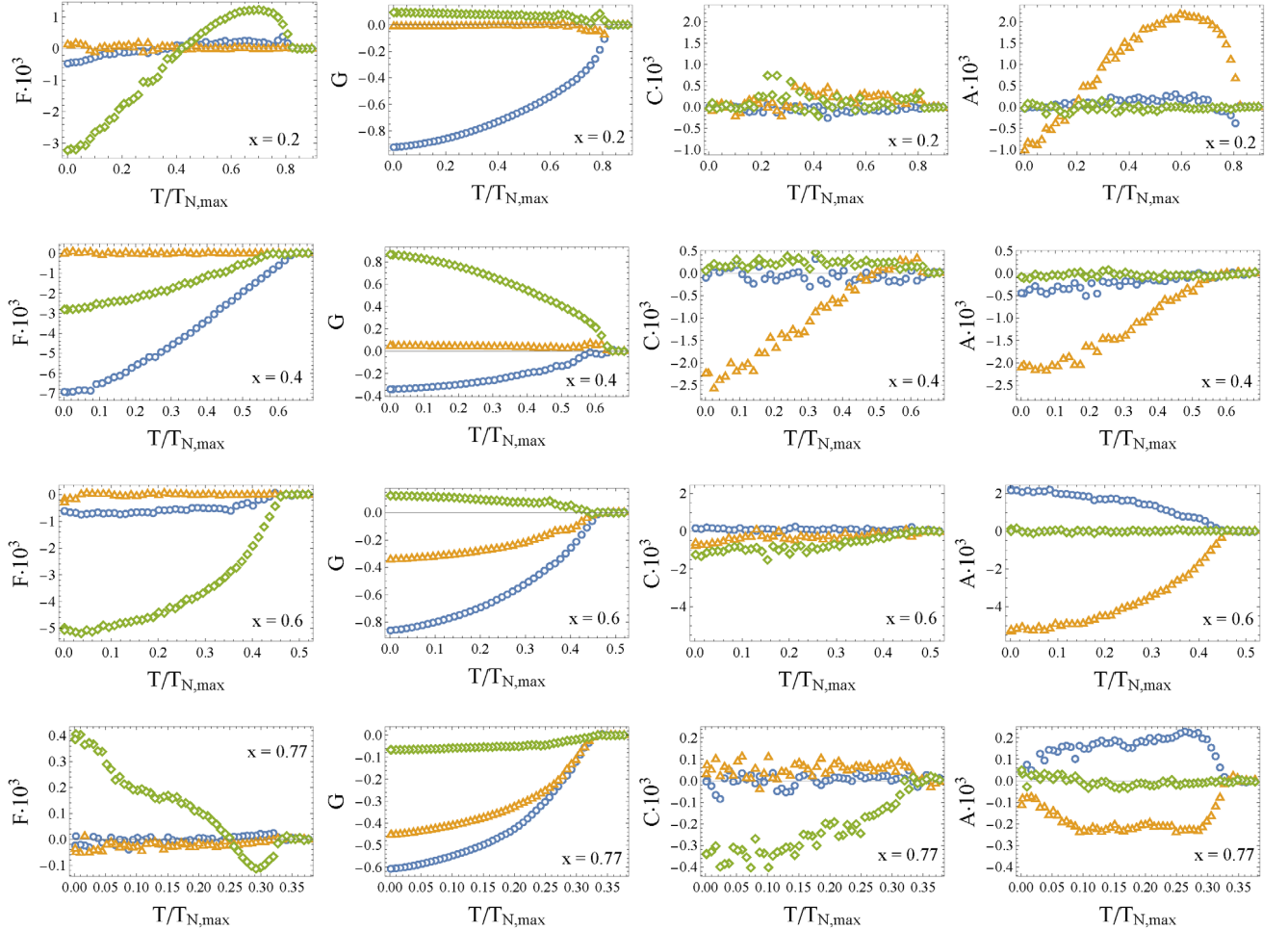

**Figure 4.** Temperature dependences of the basis vectors at different chromium concentrations, blue circles are the projection of the corresponding vector along $a$ the crystal axis, orange triangles are along $b$ the axis, green diamonds are along $c$ the axis

A characteristic feature of the method is that in the region of intermediate concentrations, for the chosen model parameters, we discover a phase $G_{xyz}$ that includes all the vector components $F, G, A, C$ (Fig. 4). Thus, there exists an angular configuration that, in the mixed composition YFe$_{1-x}$Cr$_x$O$_3$, is preferable to the "parent" one $\Gamma_4$ ($G_x, A_y, F_z$).

The MC2 method, although essentially classical, takes into account the operator nature of spin using semiclassical wave functions of the form (6) is sufficient to demonstrate the presence of phases different from $\Gamma_4$ (i.e., the possibility of spin reorientation) solely due to the competition of Dzyaloshinskii vectors without including additional mechanisms, such as external fields and single-ion anisotropy. Verification of the MC2 method in the absence of competition, with codirectional vectors $\boldsymbol{d}_{FeFe}$, $\boldsymbol{d}_{CrCr}$ and $\boldsymbol{d}_{FeCr}$, did not detect any angular configurations, only the phase is observed $\Gamma_4$ $(G_x, A_y, F_z)$.

Let us also pay attention to the issue of determining critical temperatures $T_C$, for example, when the system transitions from a disordered paramagnetic state to the phase $\Gamma_4$.

In the case of classical MC methods with the Metropolis algorithm, the temperature data turns out to be significantly lower than in the case of the mean-field approximation. On the one hand, it is known that the MFA tends to overestimate $T_C$, since such methods do not take into account important local correlations and fluctuations. On the other hand, when modeling systems with off-diagonal operators in the Hamiltonian, a strong underestimation occurs $T_C$ in classical MC methods. We attribute this to the "indestructible" dispersion of energy and order parameters caused by the continuity of the spectrum of single-site operators. When choosing a new state at a site, we are limited only by the normalization condition; thus, at any temperature, there is a new state of the site sufficiently close in energy that will, with a high probability, be accepted during an elementary step of the Metropolis algorithm. For this reason, in this work, we were interested only in the magnetic configuration of the mixed composition YFe$_{1-x}$Cr$_x$O$_3$, and the results above were presented in relative temperatures $T/T_{N,max}$, where $T_{N,max}$ is the temperature of the transition from the disordered paramagnetic state for YFeO$_3$ ($T_{N,max} = 320$ K in the MC1 method, and $T_{N,max} = 85$ K in the MC2 method). For a more accurate study of critical temperatures by the classical MC method, other algorithms are required [31,33,34,39].

## 5 Conclusion

In this work, we have developed a program for Monte Carlo simulation of magnetic moments in the mixed orthoferrite-orthochromite YFe$_{1-x}$Cr$_x$O$_3$. Key features are explained by the competition of Dzyaloshinskii vectors. It is confirmed that the system has a first compensation near $x \approx 0.2$, and a second compensation is predicted near $x \approx 0.8$. While in the parent systems YFeO$_3$ and YCrO$_3$ only the phase is observed $\Gamma_4(G_x, A_y, F_z)$, in their mixed composition an angular $G_{xyz}$ phase is observed, including all components of the basis vectors $\boldsymbol{F, G, C, A}$. The presence of the $G_{xyz}$ phase indicates the possibility of spin reorientation, but a detailed description requires additional research, consideration of single-ion spin anisotropy, and modification of the algorithm that determines the elementary step of the Monte Carlo method.

**Financing the work**

This study was supported financially by the Ministry of Science and Higher Education of the Russian Federation, Project FEUZ-2023-0017